\documentclass[aps,prl,twocolumn,showpacs,superscriptaddress,floatfix]{revtex4-1} 

\usepackage{graphicx}  
\usepackage{amsmath}
\usepackage{bm}        
\usepackage{amssymb}   
\usepackage{lipsum}
 
\usepackage{float}
\usepackage{color}

\begin{document}
\title{A colloidal viewpoint on the finite sphere packing problem: the sausage catastrophe}

\author{Susana Mar\'in-Aguilar}%
\email{s.marinaguilar@uu.nl}
\affiliation{ 
Soft Condensed Matter \& Biophysics, Debye Institute for Nanomaterials Science, Utrecht University, Princetonplein 1, 3584 CC Utrecht, The Netherlands
}%

\author{Fabrizio Camerin}%
\email{f.camerin@uu.nl}
\affiliation{ 
Soft Condensed Matter \& Biophysics, Debye Institute for Nanomaterials Science, Utrecht University, Princetonplein 1, 3584 CC Utrecht, The Netherlands
}%

\author{Stijn van der Ham}%
\affiliation{Active Soft Matter and Bio-inspired Materials Lab, Faculty of Science and Technology, University of Twente, PO Box 217, 7500 AE Enschede, The Netherlands.
}%

\author{Andréa Feasson}%
\affiliation{Active Soft Matter and Bio-inspired Materials Lab, Faculty of Science and Technology, University of Twente, PO Box 217, 7500 AE Enschede, The Netherlands.
}%
\author{Hanumantha Rao Vutukuri}%
\email{h.r.vutukuri@utwente.nl}
\affiliation{Active Soft Matter and Bio-inspired Materials Lab, Faculty of Science and Technology, University of Twente, PO Box 217, 7500 AE Enschede, The Netherlands.
}%

\author{Marjolein Dijkstra}%
\email{m.dijkstra@uu.nl}
\affiliation{ 
Soft Condensed Matter \& Biophysics, Debye Institute for Nanomaterials Science, Utrecht University, Princetonplein 1, 3584 CC Utrecht, The Netherlands
}%

\date{\today}
\begin{abstract}

It is commonly believed that the most efficient way to pack a finite number of equal-sized spheres is by arranging them tightly in a cluster. However, mathematicians have conjectured that a linear arrangement may actually result in the densest packing.
Here, our combined experimental and simulation study provides a realization of the finite sphere packing problem by studying non-close-packed arrangements of colloids in a flaccid lipid vesicle.
We map out a state diagram displaying linear, planar and cluster conformations of spheres, as well as bistable states which alternate between cluster-plate and plate-linear conformations due to  membrane fluctuations.
Finally, by systematically analyzing truncated polyhedral packings, we identify  clusters   of $56\leq N \leq 70$ spheres, excluding $N=57$ and 63, that pack more efficiently than linear arrangements.
\end{abstract}

\maketitle

The best way of packing  spheres has a long history, dating back to the works of Kepler, Gauss, and Newton, while the British sailor Raleigh was also intrigued by this problem as he searched for an efficient way to stack cannonballs on his  ship~\cite{hales2011historical}. Sphere packings also have applications in coding theory, crystallography, and in understanding  mechanical and geometrical properties of materials~\cite{cohn2016conceptual,cohn2003new,kontorovich2019geometry,torquato2010jammed,parisi2010mean}. In 1611, Kepler  conjectured that the densest  packing of an {\em infinite} number of identical, non-overlapping spheres in three dimensions is the `cannonball' stacking or the face-centered cubic (FCC) crystal, which fills space with an efficiency of $\sim 74\%$. This conjecture was  proven mathematically only very recently~\cite{hales2005proof,hales2017formal}.  

In reality, however, all packings are {\em finite}, which poses the question of what is the most efficient way to pack equal-sized  spheres in either a container with a pre-defined shape such as a sphere~\cite{manoharan2003dense,arkus2011deriving, de2015entropy},  or in a flexible container such as the smallest convex hull that can be wrapped around the spheres~\cite{maibaum2001colloids}. 
Surprisingly, the densest packing of spheres within  their convex hull is not always a compact cluster of spheres. 
On the contrary, in 1975 the mathematician Fejes-T\'oth theorized that the densest packing is always the linear arrangement where all the spheres lie on a line in dimensions $d\geq5$, and that for lower dimensions an upper limit exists where the linear conformation ceases to be the densest one. In $d=4$, this sudden transition from a linear arrangement to a cluster shape is shown to happen at a number of spheres $N=375769$, and is typically named the ``sausage catastrophe"~\cite{henk2021packings}. 
In $d=3$, the sausage conformation minimizes the volume of the convex hull for $N=57,58,63, 64$ and for $N\leq55$ particles~\cite{wills1985density,gandini1992finite}, while above this limit the densest configuration becomes  a three-dimensional cluster, thereby avoiding the plate conformation where the centers of the spheres are positioned on a plane~\cite{betke1982slices, toth2022miscellaneous}. However, the precise structure of these clusters, which are denser than the sausage, remains largely unknown.
 
Despite its fundamental importance, the finite sphere packing problem has primarily been studied mathematically, and a physical realization, even for a limited number of spheres, is still lacking. Colloidal hard spheres can serve as an ideal model system for exploring this issue~\cite{manoharan2015colloidal,kurita2010experimental}. Using their excluded-volume interactions in combination  with an appropriate flexible container provides a way to explore the different conformations that can be adopted by the spheres. Inspired by the theoretical work of Maibaum {\em et al.}~\cite{maibaum2001colloids}, we use giant unilamellar vesicles (GUVs), which are effectively large elastic containers in which colloidal particles can be enclosed~\cite{bao2021production,okano2018deformation,vutukuri2020active}. The dynamics of GUVs can be studied using  confocal microscopy, enabling direct observation of their shape fluctuations~\cite{dimovagiant,pecreaux2004refined}. GUVs have the ability to  alter their shape in response to external stimuli such as changes in osmotic pressure~\cite{pencer2001osmotically,wennerstrom2022thermal} and forces exerted by  passive~\cite{natsume2010shape} and active particles~\cite{vutukuri2020active}. 
However, experimental realizations of sausage- and plate-like arrangements are elusive.

\begin{figure*}[t!]
\begin{center}
\includegraphics[width=1.0\linewidth]{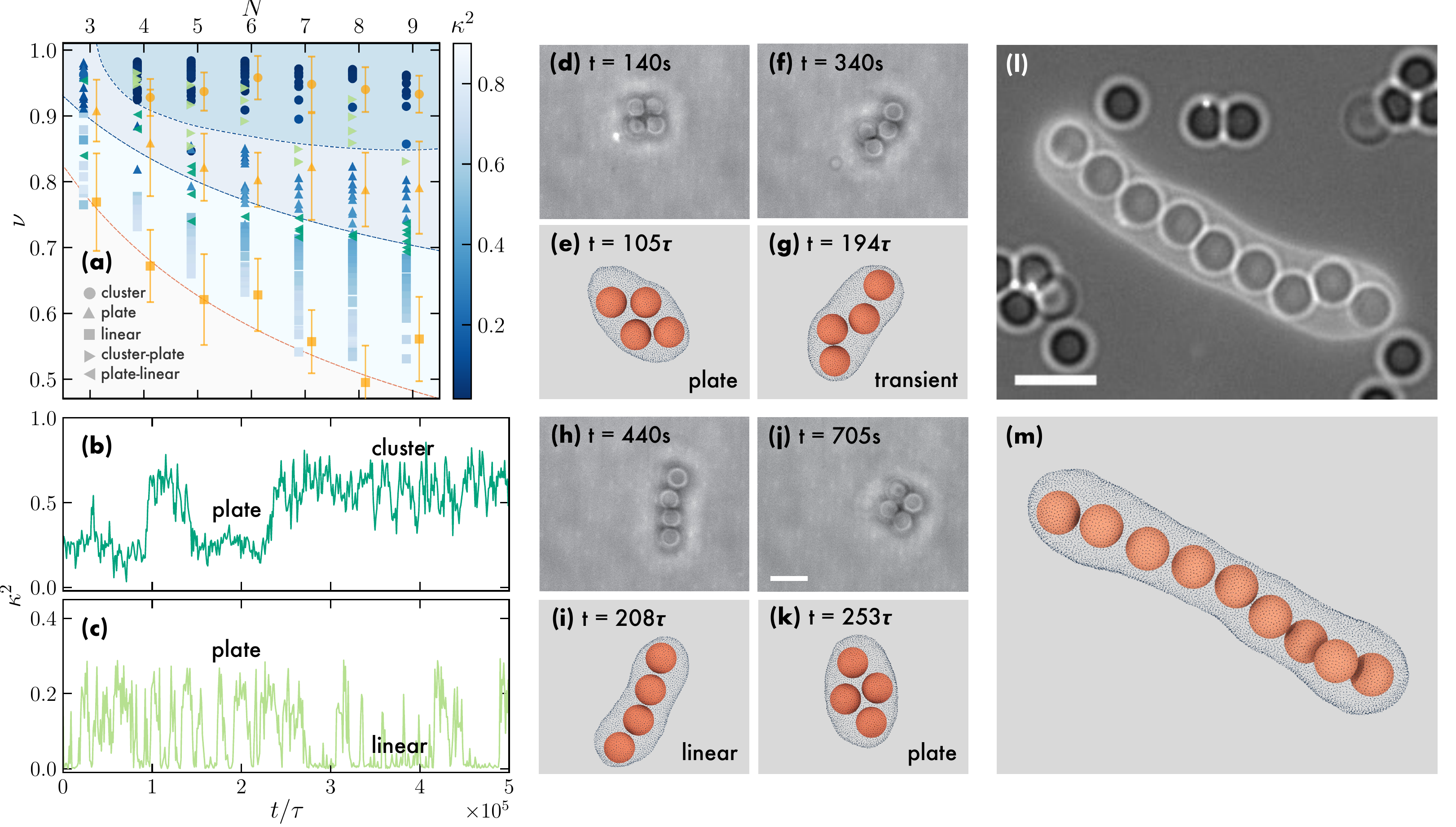}
\end{center}
\vspace{-0.5cm}
\caption{a) State diagram of colloidal hard spheres enclosed in GUVs as a function of number of colloids $N$ and reduced volume $\nu$. Different symbols denote different colloid arrangements, according to the legend. Blue-shaded symbols are numerical state points colored according to the anisotropy parameter $\kappa^2$, green symbols are for numerical bistable state points, orange symbols are for experiments. The error bars correspond to the standard deviation of the data points combined with an image analysis error (see SM~\cite{sm}). The orange line is for $\nu_{lin}(N)$, while blue lines are guides-to-the-eye for identifying the regions in which linear, planar and cluster configurations are prevalent. 
(b,c) $\kappa^2$ as a function of simulation time $t/\tau$ for a bistable line-plate and plate-cluster state, respectively. (d-k) Sequence of time-lapse images from 2D confocal and  bright field microscopy, with simulations revealing a bistable line-plate state point, and (l,m)  a $N=9$ linear arrangement. Scale bars are 5 $\mu m$.}
\label{fig:phasediagram}
\end{figure*}

In this Letter, we examine the effect of membrane fluctuations driven by Brownian colloidal spheres inside the vesicle on the different conformations of vesicle and colloids. 
In particular, we determine whether the sausage catastrophe can be observed in  systems where the spheres are not close-packed and that accounts for the entropy of the positional degrees of freedom of the colloidal particles.
Combining experiments on silica or polystyrene particles enclosed in GUVs with computer simulations, we demonstrate that colloidal spheres can form linear, plate, and cluster conformations under certain physical conditions. We summarize our results for $N\leq 9$ in a state diagram displaying the aforementioned conformations of spheres as a function of a single-order parameter.
Additionally, we discover bistable states in which the system alternates between cluster-plate and plate-linear conformations. Finally, we identify the required conditions for the formation of finite clusters with high packing for a large number of spheres and study them systematically. In this way, we uncover the sphere clusters that pack better than the sausage, and we discover clusters of $N=58$ and $64$ spheres that pack denser than the linear conformation, contradicting prior findings~\cite{wills1985density,gandini1992finite}.

We begin our study by exploring the conformations of $N$ colloids confined in a fluid vesicle both in experiments and simulations. In experiments, we use a droplet transfer method to encapsulate colloidal particles of size $2.12$ $\mu$m in GUVs ~\cite{dimovagiant,vutukuri2020active}. Next, the vesicles are exposed to hypertonic shock, where the solute concentration is larger outside than inside the vesicle, to control their morphology, see Supplemental Material (SM)~\cite{sm}. The vesicle morphologies and particle dynamics are followed by a fast confocal scanning microscope.
In the molecular dynamics simulations, different vesicle shapes are explored using a mesh-less membrane model~\cite{yuan2010dynamic,fu2017lennard}, in which the lipids are represented in a coarse-grained fashion by spheres of diameter $\sigma$ with an orientation-dependent interaction. The membrane is designed to have an approximately null surface tension, similar to the experiments. 
We use explicit solvent particles to control the initial shape of the vesicle and either exert an external pressure on the membrane or create an osmotic pressure difference between its inner and outer part. Simulations are performed with LAMMPS~\cite{thompson2022lammps}. Additional details on the simulations are described in the SM~\cite{sm}.
Once the vesicle has reached the desired shape, we insert a number of colloids $N \in [3,9]$ with a  diameter of $\sigma_c=12\sigma$ in the vesicle. The colloids interact with a repulsive Weeks-Chandler-Andersen potential. 

In both experiments and simulations, the size (and therefore the surface area) of the vesicle is adjusted depending on $N$ to study the impact of the  positional degrees of freedom of the spheres on the vesicle. We describe the state of the vesicle  by the  reduced volume $\nu$, which is the ratio between the volume of the vesicle $V_v$ and the volume of a  sphere $V_{s}$ with the same surface area as the vesicle $A_{v}$:
\begin{equation}
    \nu=\frac{V_{v}}{V_{s}}=3\sqrt{4\pi}\frac{V_{v}}{A_{v}^{3/2}}.
\end{equation}
The parameter $\nu$ takes positive values between $0<\nu\leq1$, with $\nu=1$ for an undeformed spherical vesicle~\cite{jaric1995vesicular}. 
In order to access $\nu$,   we determine in simulations the vesicle's volume and area by constructing a surface mesh around the vesicle~\cite{stukowski2009visualization}, while in experiments, they are extracted from the XYZ confocal data using an ImageJ plugin~\cite{sm}.
Each colloid conformation is characterized by the anisotropic shape parameter $\kappa^2=3(a_x^2+a_y^2+a_z^2)/2(a_x+a_y+a_z)^2-1/2$, derived from the eigenvalues $(a_x, a_y, a_z)$ of the diagonalized gyration tensor, constructed from the $x$, $y$, and $z$ coordinates of the colloids.
This proves particularly effective in distinguishing  linear conformations, characterized by  $\kappa^2 \gtrsim 0.5$, planar arrangements, indicated by  $0.2 \lesssim \kappa^2\lesssim 0.3$, and clusters with a more  isotropic shape, represented by  $\kappa^2\approx 0$.

In Fig.~\ref{fig:phasediagram}(a), we summarize our results both from experiments and simulations in a state diagram as a function of $\nu$ and number of colloids $N\in[3,9]$, where we use different symbols to denote different colloid conformations as identified by $\kappa^2$ and visual inspection. Such arrangements are showcased in Figure S1, while Fig.~\ref{fig:phasediagram}(l) and (m) show the longest linear conformation that we observed in the fluorescence and bright-field microscope and simulations, respectively (see also Movie S1).
We note that, for higher $N$, elongated vesicles exhibited excessive bending in both experiments and simulations.
The color coding denotes the values of $\kappa^2$, while the orange dashed line represents the reduced volume corresponding to the optimal linear packing of all  spheres in contact, i.e.  a spherocylinder of length $N-1$ and radius $\sigma_c/2$ for which $\nu_{lin} (N)=(4/3+2(N-1))/(4/3N^{3/2})$. We note that state points from simulations cover a wide range of packing fractions $\eta \approx 0.05-0.5$ of the colloids in the vesicle (see Fig. S2) to be compared with $\eta\simeq 0.7$ for the tightest sausage. Interestingly, we find that the reduced volume $\nu$ accounts for differences in  packing fraction, making the state diagram independent of vesicle size. For each $N$, we identify different regions denoted by different shades of blue where the linear, planar, and cluster configurations are prevalent. Upon increasing the number of colloids $N$, we find that the linear conformation becomes stable in a  wider range of $\nu$. For all $N$, we find  clusters  for $\nu > 0.9$, which are the least packed,  plate conformations for intermediate $\nu$, and linear arrangements for the lowest $\nu$ corresponding to the best-packed ones. We thus confirm experimentally and in simulations that the sausage maximizes the packing efficiency for $N\in[3,9]$.
The state diagram shows good agreement between experimental findings and simulation results.

More surprisingly, we find bistable regions due to the combined membrane and shape fluctuations driven by the colloids inside the vesicle.
These regions are identified in simulations by calculating the order parameter $\kappa^2$ as a function of time.  Fig.~\ref{fig:phasediagram}(b) and (c) demonstrate bistability between  linear-plate  and plate-cluster conformations, respectively, for $N=4$, where we can clearly see that the order parameter $\kappa^2$ fluctuates between
the values corresponding to the  different conformations.  We denote these bistable state points as left and right green triangles in the state diagram in Fig.~\ref{fig:phasediagram}(a). 
Fig.~\ref{fig:phasediagram}(d-k) show time-lapse snapshots of a bistable state point in both experiments and simulations, where we observe that a vesicle in an initial planar configuration transitions to a linear conformation (Movie S2). The transition is reversible as the colloids return to a planar conformation at longer times.

In this way, we have determined the physical conditions that allow the observation of linear, planar, and cluster conformations of $N\in [3,9]$ hard-sphere colloids in a flexible vesicle. However, the number of colloids is rather limited and significantly lower than that predicted to result in the sausage catastrophe.
We, therefore, investigate by means of simulations the possibility of observing  the sausage catastrophe in a flexible vesicle and identifying cluster conformations of spheres that pack better than the linear arrangement.  
To this end, we place $N\in[20,150]$ colloids in a spherical vesicle in such a way that it is as tight as possible without breaking the layer of beads composing the vesicle, resulting in a maximum packing fraction $\eta\approx 0.4$.
We simulate colloids in the vesicle and collect the different cluster conformations. In addition, we also consider compact minimum-energy clusters of Lennard-Jones (LJ) particles taken from a database~\cite{mravlak2016structure}. Using an optimization protocol to reach the hard-sphere limit~\cite{sm}, we determine the colloid  packing fraction as $\eta_{ch}=NV_0/V_{ch}$, where $V_0$ is the volume of each colloid and $V_{ch}$ the volume of the convex hull enclosing the colloids. By using $\eta_{ch}$, we effectively study the packing in the tightest possible container, which can directly be compared with the ideal linear packing $\nu_{lin}$. 

\begin{figure}[!t]
\begin{center}
\includegraphics[width=0.9\linewidth]{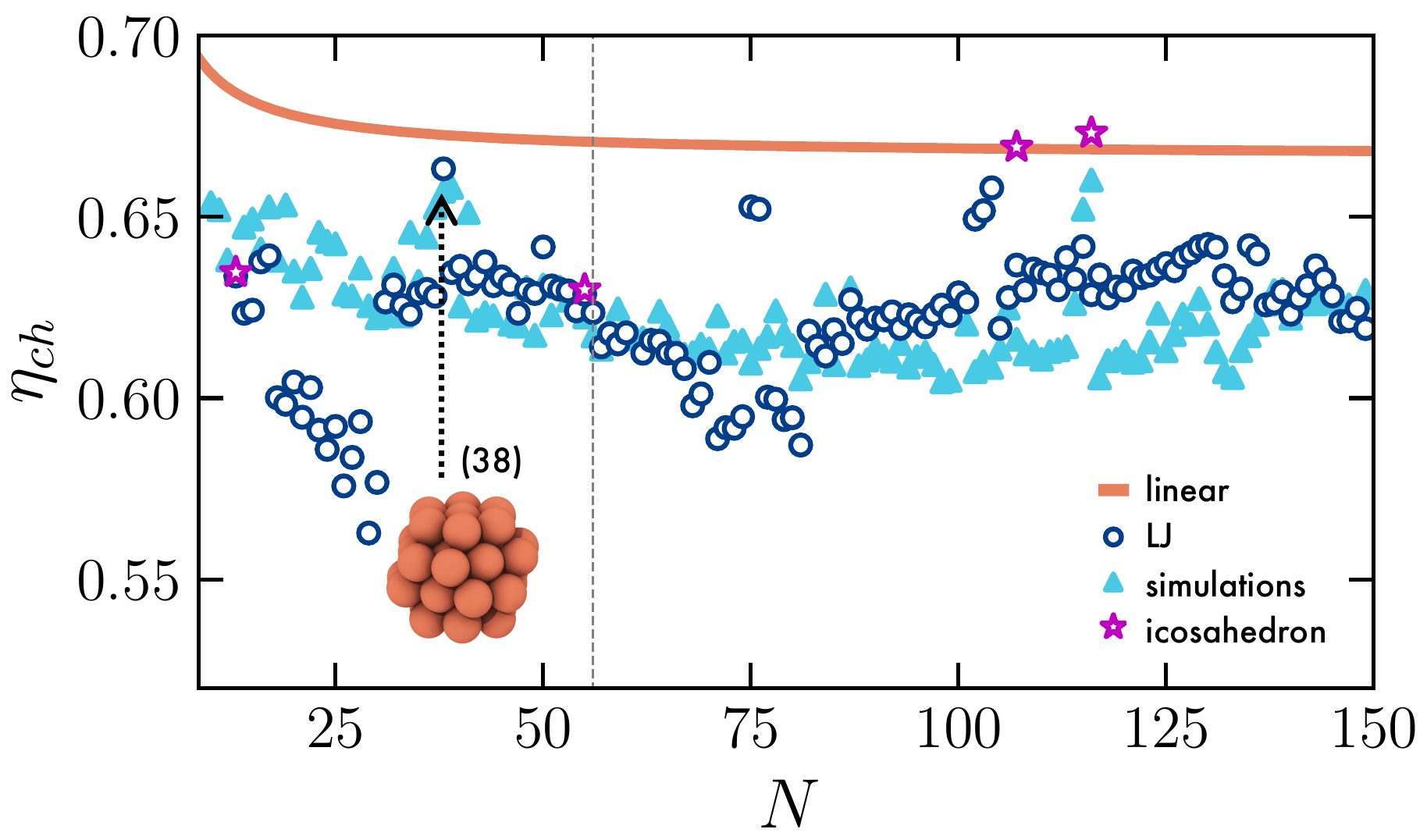}
\end{center}
\vspace{-0.5cm}
\caption{(a) Packing fraction $\eta_{ch}$ of minimum-energy Lennard-Jones (LJ) clusters, clusters obtained from simulations, and icosahedra as a function of $N$. The $N=38$ LJ cluster is also shown. 
}
\label{fig:etaN}
\end{figure}

We report $\eta_{ch}$ of simulated and minimum-energy LJ clusters in Fig.~\ref{fig:etaN} as a function of $N$. 
We indicate with a vertical dashed line the number of particles $N$ at which a cluster is expected from literature~\cite{henk2021packings,wills1985density,gandini1992finite} to have a higher $\eta_{ch}$ than the linear conformation, that is where the sausage catastrophe occurs. For comparison, we also plot $\eta_{ch}$ of icosahedra for various $N$, as these structures are expected to pack locally very efficiently~\cite{de2015entropy,wang2018magic}.  We observe that all the clusters studied have a lower $\eta_{ch}$ than the linear conformation, with the exception of two icosahedra with $N>100$, which have slightly higher packing fractions than the linear conformation. Additionally, we find that the clusters obtained from simulations generally have a similar $\eta_{ch}$ of the LJ clusters.
However, it is worth noting that only a few clusters exhibit significantly higher packing fractions than the others, thus approaching linear packing. These clusters have higher bond-orientational order parameter values $q_6$~\cite{steinhardt1983bond} compared to the other clusters (see Fig. S7), which points towards an underlying FCC crystalline order in their arrangement. 
As an example, we focus on the LJ cluster with $N=38$ (see Figs.~\ref{fig:etaN} and S8), which was previously found to represent a minimum-energy configuration of the LJ potential and to be stable for a wide range of LJ parameters~\cite{doye1997structural}. The cluster has a truncated octahedral shape based on an FCC arrangement, which allows for the presence of regular two-dimensional patterns on its surfaces.

Based on these findings, we expect that clusters that surpass the linear packing will exhibit similar characteristics to the cluster just analyzed. More precisely, they should retain an FCC structure. Therefore, in the subsequent part of our study, we examine a number of ordered arrangements in which spheres are in contact. 

\begin{figure}[!b]
\begin{center}
\includegraphics[width=1\linewidth]{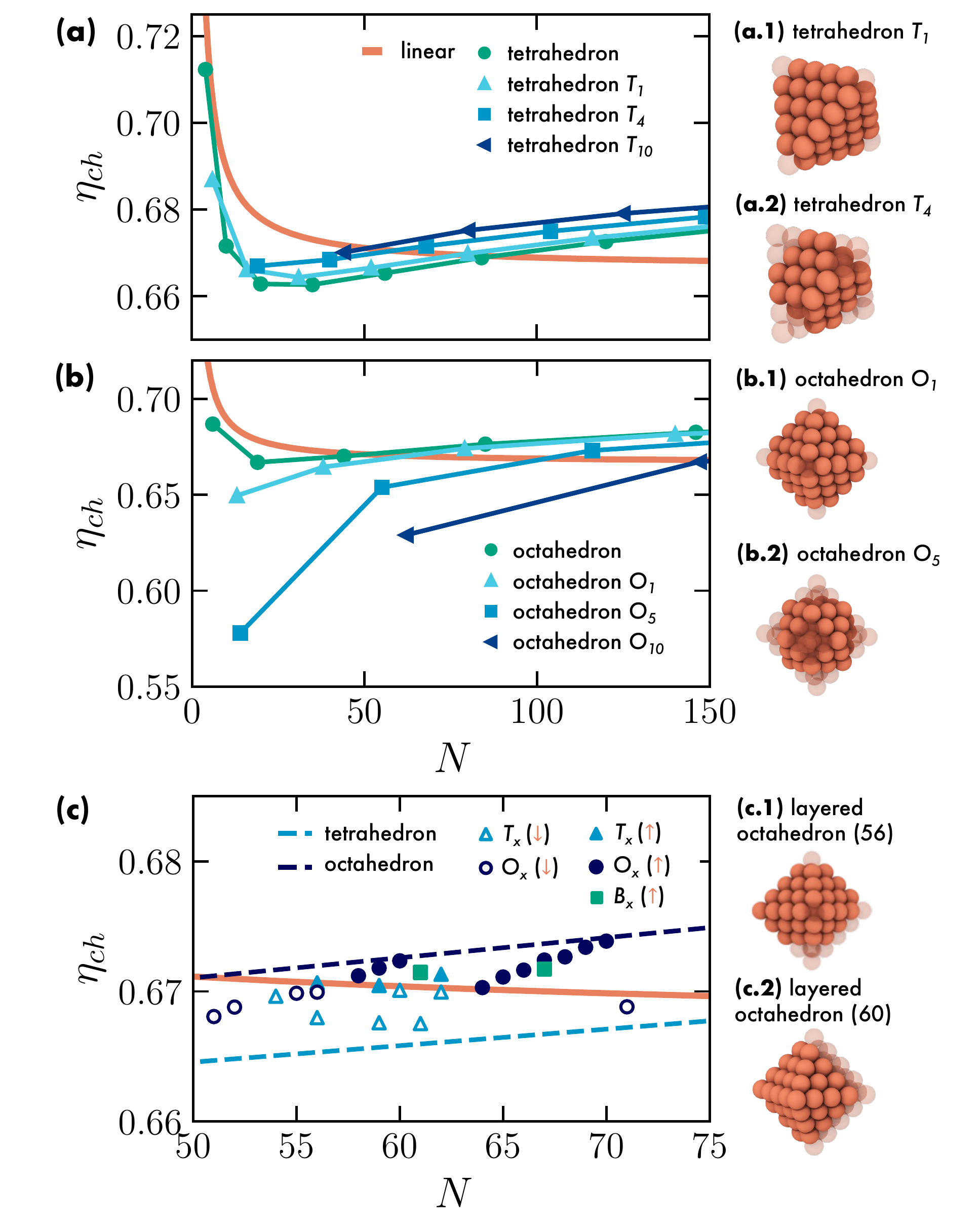}
\end{center}
\vspace{-0.5cm}
\caption{
Packing fraction $\eta_{ch}$ of clusters of spheres in their convex hull as a function of $N$ for the linear arrangement of spheres (orange line) and for (a) regular and regularly truncated tetrahedra, (b) regular and regularly truncated octahedra and (c) for a number of irregularly truncated tetrahedra, octahedra and bipyramyds. In (c), open symbols are for irregularly truncated tetrahedra and octahedra for which $\eta_{ch}<\eta_{ch}^{linear}$ ($\downarrow$); closed symbols are used if $\eta_{ch}>\eta_{ch}^{linear}$ ($\uparrow$). 
Each panel is accompanied by example clusters, where the original non-truncated configuration is shown with transparent spheres.}
\label{fig:saus}
\end{figure}

We confirm that two-dimensional structures' packing is always less favorable than linear arrangements (see Fig. S9). This is consistent with previous work, which demonstrated that the optimal packing can only be achieved in either linear or cluster arrangements~\cite{betke1982slices}.
Focusing on three-dimensional clusters, we start with some of the simplest polyhedra such as tetrahedra, octahedra and bipyramids. To explore a large variety of clusters, we consider polyhedra of different sizes constructed by slicing a close-packed FCC crystal, and perform subsequent cuts to the vertices of all shapes. We use the notation $X^k_n$ ($X_n$ for isotropic cuts), where $X=T,O,B$ for tetrahedra, octahedra or bipyramid, respectively, $n$ represents the number of particles removed from each vertex, and $k$ denotes the number of vertices from which particles are removed.
In Fig.~\ref{fig:saus}(a), we show the convex-hull packing fraction $\eta_{ch}$ and typical configurations for some sliced tetrahedra $T_1$ and $T_4$.
For $N=4$, $\eta_{ch}$ of the regular tetrahedron is very close to that of four spheres on a line. Increasing $N$ causes $\eta_{ch}$ to decrease until it reaches a minimum, for then crossing the packing fraction of the line at $N=84$. 
Similarly, Fig.~\ref{fig:saus}(b) reports $\eta_{ch}$ for regularly truncated octahedra with corresponding configurations.
Contrary to tetrahedra, octahedra show a decrease in $\eta_{ch}$ as more particles are removed, and the crossover to the packing fraction of a sausage occurs at larger $N$. In general, $\eta_{ch}$ for octahedra is lower than that of tetrahedra, particularly for $N < 50$.

Next, we analyze how the packing is affected by an asymmetric removal of spheres from regular tetrahedra, octahedra and bipyramids, i.e. by removing a different number of spheres from each vertex or layer of the polyhedra. Specifically, we focus on the region close to the $N$ where the sausage catastrophe was predicted to occur. We present the results in Fig.~\ref{fig:saus}(c), with filled symbols denoting tetrahedra, octahedra and bypiramids whose packing fraction exceeds that of the linear arrangement (see Figs. S10-S12 and interactive HTML files~\cite{sm}).
While the mathematical requirements for tetrahedra and bipyramids to form clusters denser than the sausage have been established~\cite{gandini1992finite}, such a prediction has not been made for octahedra. Remarkably, we observe that in this range of $N$ only polyhedral clusters that are sliced asymmetrically pack denser than the sausage.
Our results indicate that regular or regularly cut polyhedra are generally not the arrangements that maximize packing efficiency.
Furthermore, we observe no specific correlations between different structural elements of the clusters that can directly influence packing, such as the number of faces, edges, or vertices (see Table S4). It appears that better packing results from nontrivial combinations of all these elements, with each contributing marginally to minimize the available volume.

In conclusion, our study sheds new light on the finite sphere packing problem and provides valuable insights on the most efficient methods for packing spheres in a closed, flexible container. Our simulation predictions are consistent with the experimental observations. We demonstrate that a low-tension vesicle is an ideal system for studying linear, planar, and cluster configurations formed by colloidal hard spheres.
By constructing a general state diagram based on a single-order parameter describing the reduced volume of the vesicle, we can differentiate between stable and bistable conformations. Our simulations reveal that simply packing spheres in a vesicle does not yield configurations with higher packing fractions than the linear one, even with  higher number of particles. However, by examining the structure and arrangement of the particles, we find that higher packing fractions can only be achieved with faceted ordered clusters, such as with truncated tetrahedra and octahedra in the region where the sausage catastrophe emerges. Although these clusters cannot be realized by packing spheres in a flexible vesicle, they may be experimentally observable in systems with strong attractive interactions, such as gold nanoparticles and platinum clusters~\cite{schmid1992large, eguchi2012simple, mori2016determining,tlahuice2019new}.

\bigskip

\begin{acknowledgements}
S.M.-A., F.C., and M.D. thank Gerardo Campos-Villalobos and Rodolfo Subert for early contributions to this work and for useful discussions. H.R.V. and S.v.d.H. thank Frieder Mugele and Mireille Claessens for kindly providing access to confocal and fluorescence microscopes.

S.M.-A. and F.C. contributed equally to this work. S.M.-A. and F.C. performed the numerical simulations and analysis.
S.v.d.H. and A.F. performed the experiments. S.v.d.H. performed the experimental analysis. H.R.V. supervised S.v.d.H. and A.F., H.R.V. and M.D. conceived and designed the project. S.M.-A., F.C., H.R.V., and M.D. wrote the original draft of the paper. All authors participated in the discussions, and reviewed and edited the manuscript.

S.M.-A., F.C. and M.D. acknowledge funding from the European Research Council (ERC) under the European Union's Horizon 2020 research and innovation program (Grant agreement No. ERC-2019-ADG 884902, SoftML).
\end{acknowledgements}

\bibliographystyle{apsrev4-1}
\bibliography{vesicle}
\end{document}